\newif\ifsubmode
\submodefalse

\ifsubmode
  \documentclass[12pt,preprint]{aastex}
  \received{}
  \revised{}
  \accepted{}
\else
  \documentclass{emulateapj}

  \usepackage{apjfonts}
  \usepackage{mathptmx}
  \usepackage{graphics}
  \usepackage{graphicx}
  \slugcomment{Accepted to the Astrophysical Journal Supplement Series}

\fi
\usepackage{epsf}

\newcommand{\tabletwo}{
\tablewidth{310pt}
\tablecaption{Measured \spitzer\ 24~$\mu\mathrm{m}$ Number
  Counts\label{table:counts}}
\tablehead{ 
\colhead{Log $S_\nu$} & \colhead{$dN/dS_\nu$} & \colhead{$\delta( dN/dS_\nu )$}  & 
\colhead{Log $S_\nu$} & \colhead{$N(>S_\nu)$} & \colhead{$\delta( N(>S_\nu) )$} \\
\colhead{(mJy)} & \colhead{(mJy$^{-1}$ sr$^{-1}$)} & \colhead{(mJy$^{-1}$
sr$^{-1}$)}  & 
\colhead{(mJy)} & \colhead{(sr$^{-1}$)} & \colhead{(sr$^{-1}$)}\\
  \colhead{(1)} & \colhead{(2)} & \colhead{(3)} & 
  \colhead{(4)} & \colhead{(5)} & \colhead{(6)}}
\startdata
$-1.475$ & 2.5 (1.4) $\times 10^9$ & 4.7 (1.5) $\times 10^8$ &
$-1.550$ & 2.1 (1.1) $\times 10^8$ & 1.2 (0.60) $\times 10^7$ \\
$-1.325$ & 1.5 (1.2) $\times 10^9$ & 2.7 (0.97) $\times 10^8$ &
$-1.400$ & 1.2 (0.92) $\times 10^8$ & 7.6 (4.9) $\times 10^6$ \\
$-1.175$ & 8.7 (6.9) $\times 10^8$ & 1.4 (1.4) $\times 10^8$ &
$-1.250$  & 9.6 (7.2) $\times 10^7$ & 3.7 (0.68) $\times 10^6$ \\
$-1.025$ & 5.5 (5.2) $\times 10^8$ & 5.8 (1.6) $\times 10^7$ &
$-1.100$  & 6.1 (5.6) $\times 10^7$ & 5.9 (5.6) $\times 10^5$ \\
$-0.875$ & 3.1 (3.1) $\times 10^8$ & 7.1 (4.9) $\times 10^6$ &
$-0.950$  & 3.9 (3.9) $\times 10^7$ & 4.0 (4.0) $\times 10^5$  \\
$-0.725$ & 1.7 (1.6) $\times 10^8$ & 7.6 (6.0) $\times 10^6$ &
$-0.800$  & 2.9 (2.5) $\times 10^7$ & 5.5 (5.0) $\times 10^5$ \\
$-0.575$ & 8.3 (7.5) $\times 10^7$ & 1.5 (0.66) $\times 10^6$ &
$-0.650$ & 1.8 (1.6) $\times 10^7$ & 1.8 (1.6) $\times 10^5$ \\
$-0.425$ & $3.2\times 10^7$ & $5.7\times 10^5$ & $-0.500$ & $8.6\times 10^6$ & $1.2\times 10^5$ \\
$-0.275$ & $1.2\times 10^7$ & $1.9\times 10^5$  & $-0.350$ & $4.4\times 10^6$ & $6.2\times 10^4$  \\
$-0.125$ & $4.1\times 10^6$ & $9.6\times 10^4$  & $-0.200$ & $2.3\times 10^6$ & $3.8\times 10^4$  \\
\phs$0.025$ & $1.4\times 10^6$ & $3.9\times 10^4$  & $-0.050$ & $1.2\times 10^6$ & $2.3\times 10^4$  \\
\phs$0.175$ & $5.8\times 10^5$ & $2.1\times 10^4$  & \phs$0.100$ & $7.1\times 10^5$ & $1.7\times 10^4$  \\
\phs$0.325$ & $2.4\times 10^5$ & $1.1\times 10^4$  & \phs$0.250$ & $4.1\times 10^5$ & $1.4\times 10^4$  \\
\phs$0.475$ & $1.0\times 10^5$ & $6.5\times 10^3$  & \phs$0.400$ & $2.3\times 10^5$ & $1.0\times 10^4$  \\
\phs$0.625$ & $3.4\times 10^4$ & $2.8\times 10^3$  & \phs$0.550$ & $1.3\times 10^5$ & $7.1\times 10^3$  \\
\phs$0.775$ & $1.5\times 10^4$ & $1.6\times 10^3$  & \phs$0.700$ & $7.7\times 10^4$ & $5.5\times 10^3$  \\
\phs$0.925$ & $6.0\times 10^3$ & $8.6\times 10^2$  & \phs$0.850$ & $5.2\times 10^4$ & $7.4\times 10^3$  \\
\phs$1.075$ & $2.7\times 10^3$ & $4.9\times 10^2$  & \phs$1.000$ & $3.4\times 10^4$ & $6.0\times 10^3$  \\
\phs$1.225$ & $1.2\times 10^3$ & $2.7\times 10^2$  & \phs$1.150$ & $2.0\times 10^4$ & $4.0\times 10^3$  \\
\phs$1.375$ & $6.2\times 10^2$ & $1.7\times 10^2$  & \phs$1.300$ & $1.3\times 10^4$ & $2.8\times 10^3$  \\
\phs$1.525$ & $2.2\times 10^2$ & $8.3\times 10^1$  & \phs$1.450$ & $7.3\times 10^3$ & $2.1\times 10^3$  \\
\phs$1.675$ & $1.1\times 10^2$ & $5.0\times 10^1$  & \phs$1.600$ & $4.4\times 10^3$ & $1.5\times 10^3$ 
\enddata
\tablecomments{Col.\ (1) Flux density for differential number counts;
(2) Corrected differential counts and (3) uncertainty; (4)
Flux density for cumulative number counts;
(5) Corrected cumulative counts and (6) uncertainty. Numbers in
parentheses give the uncorrected values.}
}

\newcommand{\tableone}{
\tablewidth{0pt}
\tablecaption{Properties of Deep, \spitzer\
  Fields\label{table:fields}}
\tablehead{ \colhead{} & \colhead{R.A. } & \colhead{Decl.} & 
\colhead{$\langle I_\nu \rangle$} & \colhead{Area} & 
\colhead{$\langle t_\mathrm{exp} \rangle$} & \colhead{$C_{80\%}$} &
\colhead{$N(>C_{80\%})$}  & \colhead{$N(>300\ \mu\mathrm{Jy})$} \\
\colhead{Field} & \colhead{(J2000.0)} & \colhead{(J2000.0)} & 
\colhead{ (MJy sr$^{-1}$) } & \colhead{ (arcmin$^2$) } & 
\colhead{(sec)} & \colhead{ ($\mu$Jy)} & \colhead{ (arcmin$^{-2}$)} &
\colhead{ (arcmin$^{-2}$)}
\\
\colhead{(1)} & \colhead{(2)} & \colhead{(3)} & \colhead{(4)} &
\colhead{(5)} & \colhead{(6)} & \colhead{(7)} & \colhead{(8)} & 
\colhead{(9)}}
\startdata
Marano   & $03\ 13\ 52$  &
$-55\ 15\ 23$ & 19.7  &    \phn 1296 & \phn 236 &
170 & 2.0 & 0.9 \\
CDF--S   & $03\ 32\ 28$  &
$-27\ 48\ 30$ & 22.5  &    \phn 2092 & 1378 & \phn
83 & 4.5 & 0.7 \\
EGS  & $14\ 16\ 00$  & $+52\ 48\ 50$ &
19.5  &    \phn 1466 & \phn 450 & 110 & 3.4 & 0.7 \\
Bo\"otes & $14\ 32\ 06$  &
$+34\ 16\ 48$ & 22.7  &        32457 & \phn\phn 87 &
270 & 1.0 & 0.8 \\
ELAIS    & $16\ 09\ 52$  &
$+54\ 55\ 00$ & 18.2  & \phn\phn 130 & 3232     &
\phn 61 & 5.7 & 0.6
\enddata
\tablecomments{Col.\ (1) Field name;
(2) right ascension in units of hours, minutes, and seconds; (3)
declination in units of degrees, arcminutes, and arcseconds; (4)
mean 24~\micron\ background; (5) areal coverage; (6) mean exposure time;
(7) 80\% completeness limit; (8) source density with $S_\nu >
C_{80\%}$. (9) source density with $S_\nu > 300$~$\mu$Jy.}  }

\newcommand{\figonecap}{
  Reliability of source detection and photometry for the CDF--S, one
  of the deep \spitzer\ fields.  The \textit{thick} line shows the
  fraction of artificial sources detected in the image as a function
  of input flux density.  For the CDF--S this fraction is 80\%
  complete at $C_{80\%} = 83$~$\mu$Jy (indicated by the
  \textit{dotted} line).  The \textit{red} line shows the ratio of the
  number of sources detected in a `negative' of the 24\micron\ image
  to the number of `positive' sources as a function of flux density,
  which provides an estimate of the number of spurious sources due to
  noise features. \label{fig:completeness} }

\newcommand{\figtwocap}{
Cumulative (\textit{Left} panel) and differential (\textit{Right}
panel) 24\micron\ number counts.  The differential counts have been
normalized to a Euclidean slope,   $dN/dS_\nu \sim S_\nu^{-2.5}$.  The
solid stars show the average counts from all the \spitzer\ fields (see
Table~\ref{table:fields}), and corrected for completeness to their
respective 80\% limits.  The open star corresponds to counts brighter
than the 50\% completeness limit from the ELAIS field.  The error bars
correspond to counting uncertainties and a cosmic--variance estimate
based on the standard deviation of the field--to--field counts from
the different fields.   Each flux bin is $\Delta( \log S_\nu) =
0.15$~dex.  The shaded diamonds correspond to \textit{IRAS}
25~\micron\ number counts from \citet[and adjusted assuming $\nu_{24}
S_{\nu}(24\micron) = \nu_{25} S_\nu(25\micron)$]{hac91}.  The curves
show the predictions from various contemporary models from the
literature (see figure inset; and adjusted slightly to match the
observed \textit{IRAS} counts), and a model based on the local \iso\
15\micron\ luminosity function and assuming non--evolving galaxy
SEDs. \label{fig:counts} }

\newcommand{\spitzer}{\textit{Spitzer}}

\newcommand{\iso}{\textit{ISO}}

\newcommand{\lsim}{\lesssim}
\newcommand{\gsim}{\gtrsim}
\newcommand{\lstar}{\hbox{$L^\ast$}}

\newcommand{\etal}{et al.}
\newcommand{\eg}{e.g.}

\newcommand{\lsol}{\hbox{$L_\odot$}}

\shorttitle{THE 24\micron\ SOURCE COUNTS IN DEEP \textit{SPITZER} SURVEYS}
\shortauthors{PAPOVICH ET AL.}

\begin{document}

\title{THE 24\micron\ SOURCE COUNTS IN DEEP \textit{SPITZER}
  SURVEYS\altaffilmark{1}}
\author{\sc C.~ Papovich\altaffilmark{2}, 
  H.~Dole\altaffilmark{3}, 
  E.~Egami\altaffilmark{2}, 
  E.~Le Floc'h\altaffilmark{2},
  P.~G.~P\'erez--Gonz\'alez\altaffilmark{2},
  A.~Alonso--Herrero\altaffilmark{2,4},
  L.~Bai\altaffilmark{2},
  C.~A.~Beichman\altaffilmark{5},
  M.~Blaylock\altaffilmark{2},
  C.~W.~Engelbracht\altaffilmark{2},
  K.~D.~Gordon\altaffilmark{2},
  D.~C.~Hines\altaffilmark{2,6},
  K.~A.~Misselt\altaffilmark{2}, 
  \ifsubmode \else \\ \fi
  J.~E.~Morrison\altaffilmark{2},
  J.~Mould\altaffilmark{7}, 
  J.~Muzerolle\altaffilmark{2},
  G.~Neugebauer\altaffilmark{2},
  P.~L.~Richards\altaffilmark{8},
  G.~H.~Rieke\altaffilmark{2},
  \ifsubmode \else \\ \fi
  M.~J.~Rieke\altaffilmark{2},
  J.~R.~Rigby\altaffilmark{2}, 
  K.~Y.~L.~Su\altaffilmark{2},
  and E.~T.~Young\altaffilmark{2}
}
\altaffiltext{1}{This work is based on observations made with the Spitzer Space Telescope, which is operated by the Jet Propulsion Laboratory, California Institute of Technology under NASA contract 1407.}
\altaffiltext{2}{Steward Observatory, The University
of Arizona,  933 North Cherry Avenue, Tucson, AZ 85721; Electronic Address: papovich@as.arizona.edu}
\altaffiltext{3}{Institut d'Astrophysique Spatiale, bat 121,
  Universit\'e Paris Sud, F91405 Orsay Cedex, France}
\altaffiltext{4}{Departamento de Astrof\'{\i}sica Molecular
e Infrarroja, IEM, CSIC,
Serrano 113b, 28006 Madrid, Spain}
\altaffiltext{5}{Michelson Science Center, California Institute of
Technology, Pasadena, CA 91109}
\altaffiltext{6}{Space Science Institute, 4750 Walnut, Suite 205,
Boulder, CO 80301}
\altaffiltext{7}{National Optical Astronomy Observatory, 950
  North Cherry Avenue, Tucson, AZ 85719}
\altaffiltext{8}{Department of Physics, University of California,
  Berkeley, CA 94720}

\begin{abstract}

Galaxy source counts in the infrared provide strong constraints on the
evolution of the bolometric energy output from distant galaxy
populations.  We present the results from deep 24\micron\ imaging from
\spitzer\ surveys, which include $\approx 5\times 10^4$ sources
to an 80\% completeness of $\simeq 60$~$\mu$Jy.  The
24\micron\ counts rapidly rise at near--Euclidean rates down to
5~mJy, increase with a super--Euclidean rate between $0.4-4$~mJy, and converge below $\sim
0.3$~mJy.  The 24~\micron\ counts exceed expectations from
non--evolving models by a factor $\gsim 10$  at $S_\nu \sim 0.1$~mJy.
The peak in the differential number counts corresponds to a population
of faint sources that is not expected from predictions based on
15$\mu$m counts from \textit{ISO}.  We argue that
this implies the existence of a previously undetected population of
infrared--luminous galaxies at $z\sim 1-3$.  Integrating the counts to
$60$~$\mu$Jy, we derive a lower limit on the 24\micron\ background
intensity of $1.9\pm0.6$~nW m$^{-2}$ sr$^{-1}$ of which the majority
($\sim 60$\%) stems from sources fainter than 0.4~mJy.  Extrapolating
to fainter flux densities, sources below $60$~$\mu$Jy contribute
$0.8^{+0.9}_{-0.4}$~nW m$^{-2}$ sr$^{-1}$ to the background, which provides
an estimate of the total 24~\micron\ background of $2.7^{+1.1}_{-0.7}$~nW
m$^{-2}$ sr$^{-1}$.

\end{abstract}
 
\keywords{
cosmology: observations --- 
galaxies: evolution --- 
galaxies: high--redshift ---
galaxies: photometry ---
infrared: galaxies
}


\section{Introduction}

From the first detections of infrared (IR) luminous galaxies, it was
clear that they represent phenomena not
prominent in optically selected galaxy surveys (\eg, Rieke \& Low
1972; Soifer, Neugebauer, \& Houck 1987).   
Locally, galaxies radiate most of their emission at UV and optical
wavelengths, and only about one--third at IR wavelengths
($5-1000$~\micron; Soifer \& Neugebauer 1991).  IR number counts from
\textit{ISO} indicate that the IR--luminous sources have evolved
rapidly, significantly faster than has been deduced from optical
surveys, which implies that IR--luminous galaxies make a substantial
contribution to the cosmic star--formation rate density (e.g., Elbaz
et al.\ 1999; Franceschini et al.\ 2001).  
The detection of the cosmic background by
\textit{COBE} at IR wavelengths shows that the
total far--IR emission of galaxies in the early Universe is greater
than that at optical and UV wavelengths
\citep{fix98,hau98,mad00,fra01}, which suggests that a large
fraction of stars have formed in IR--luminous phases of galaxy
activity \citep{elb02}.  Studies from \iso\ at 15~\micron\ and
170~\micron\ have inferred that the bulk of this background originates
in discrete sources with $z\lsim 1.2$ \citep{dol01,elb02}.  However,
the peak in the cosmic IR background extends to $\sim 200$\micron\
\citep{fix98,hau01}.  The spectral--energy distributions (SEDs) of
luminous IR galaxies ($L_{\mathrm{IR}} \sim 10^{11-12}$~\lsol) peak between
$50-80$~\micron\ \citep{dal01}. If these objects
constitute a major component of the cosmic IR background, then it follows
there is a significant population of IR--luminous galaxies at $z\sim 1.5-3$ ---
distances largely unexplored by \textit{ISO}.

\ifsubmode
\else
\begin{deluxetable*}{lcccccccc}[bh]
\tableone
\end{deluxetable*}
\fi

The mid--IR 24~\micron\ band on the Multiband Imaging Photometer for
\spitzer\ \citep[MIPS,][]{rie04} is particularly well suited for
studying distant IR--luminous galaxies.   Locally,
the mid--IR emission from galaxies relates almost linearly with the
total IR luminosity over a range of galaxy types
\citep[\eg,][]{spi95,cha01,rou01,pap02}, and there are indications
this holds at higher redshifts \citep[\eg,][]{elb02}.
Because the angular resolution of \spitzer\ is significantly higher
for the 24~\micron\ band relative to 70 and 160~\micron, the
24~\micron\ confusion limit lies at fainter flux densities. This
allows us to probe the IR emission from many more sources and at
higher redshifts than with the MIPS longer wavelength bands
\citep[\eg,][]{pap02, dol03}.  Here, we present the number counts of
$\approx 5\times 10^4$ sources detected at 24~\micron\ in deep
\spitzer\ surveys, and we suggest that the faint
\spitzer\ detections probe a previously undetected population of very
luminous galaxies at high redshifts.  
Where applicable, we assume $\Omega_m
= 0.3$, $\Omega_\Lambda = 0.7$, and $H_0 = 70$~km s$^{-1}$ Mpc$^{-1}$. 


\section{The Data and Source Samples\label{section:data}}

The data used in this work were obtained in five fields from early
\spitzer\ characterization observations and from time allocated to the
MIPS Guaranteed Time Observers (GTOs).
Table~\ref{table:fields} lists their properties and
24~\micron\ source densities.
The MIPS fields used here subtend the largest areas ($\simeq
10.5$~deg$^2$) and widest range in flux density ($50-0.06$~mJy)
available to date from the \spitzer\ mission, and therefore are the
premier dataset for studying the mid--IR source counts.  

The MIPS 24~\micron\ images were processed with the MIPS GTO data analysis
tool \citep{gor04}.  The measured 
count rates are corrected for dark current, cosmic--rays, and flux
nonlinearities, and then divided by flat--fields for each
MIPS scan--mirror position.  Images are then corrected
for geometric distortion and mosaicked.  The final
mosaics have a pixel scale of $\simeq 1\farcs25$~pix$^{-1}$, with a
point--spread function (PSF) full--width at half maximum (FWHM) of $\simeq
6\arcsec$.

\ifsubmode
\else
\begin{figure}[b]
\epsscale{1.1}
\plotone{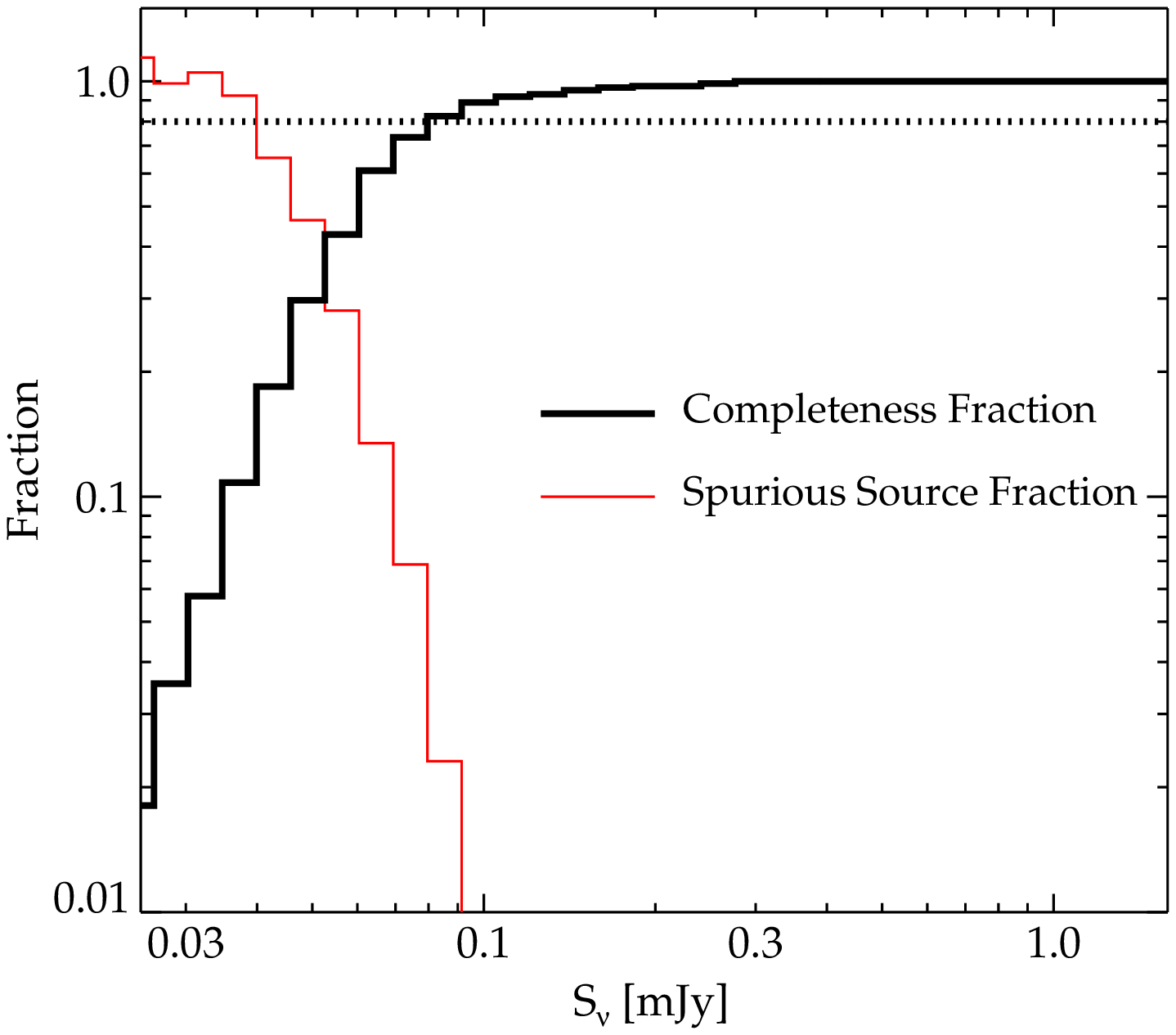}
\epsscale{1.0}
\figcaption{ \figonecap }
\end{figure}
\fi

We performed source detection and photometry using a set of tools and
simulations.
Briefly, we first subtract the image
background using a median filter roughly four times the size of source
apertures (see below), and use a version of the
DAOPHOT software \citep{ste87} to detect point sources.  We filter
each image with a Gaussian with a FWHM that is equal to that of the
MIPS 24~\micron\ PSF, and identify positive features in
10\arcsec--diameter apertures above some noise threshold. We then
construct empirical PSFs using $20-30$ bright sources in each image,
and optimally measure photometry by simultaneously fitting the
empirical PSFs to all sources within $\simeq 20$\arcsec\ of nearby
object centroids.   
The source photometry corresponds to the flux of these PSF--apertures
within a diameter of  $37\farcs4$, and we apply a multiplicative
correction of 1.14 to account for light lost outside these apertures.

\ifsubmode
\else
\begin{figure*}[th]
\epsscale{1.111}
\plottwo{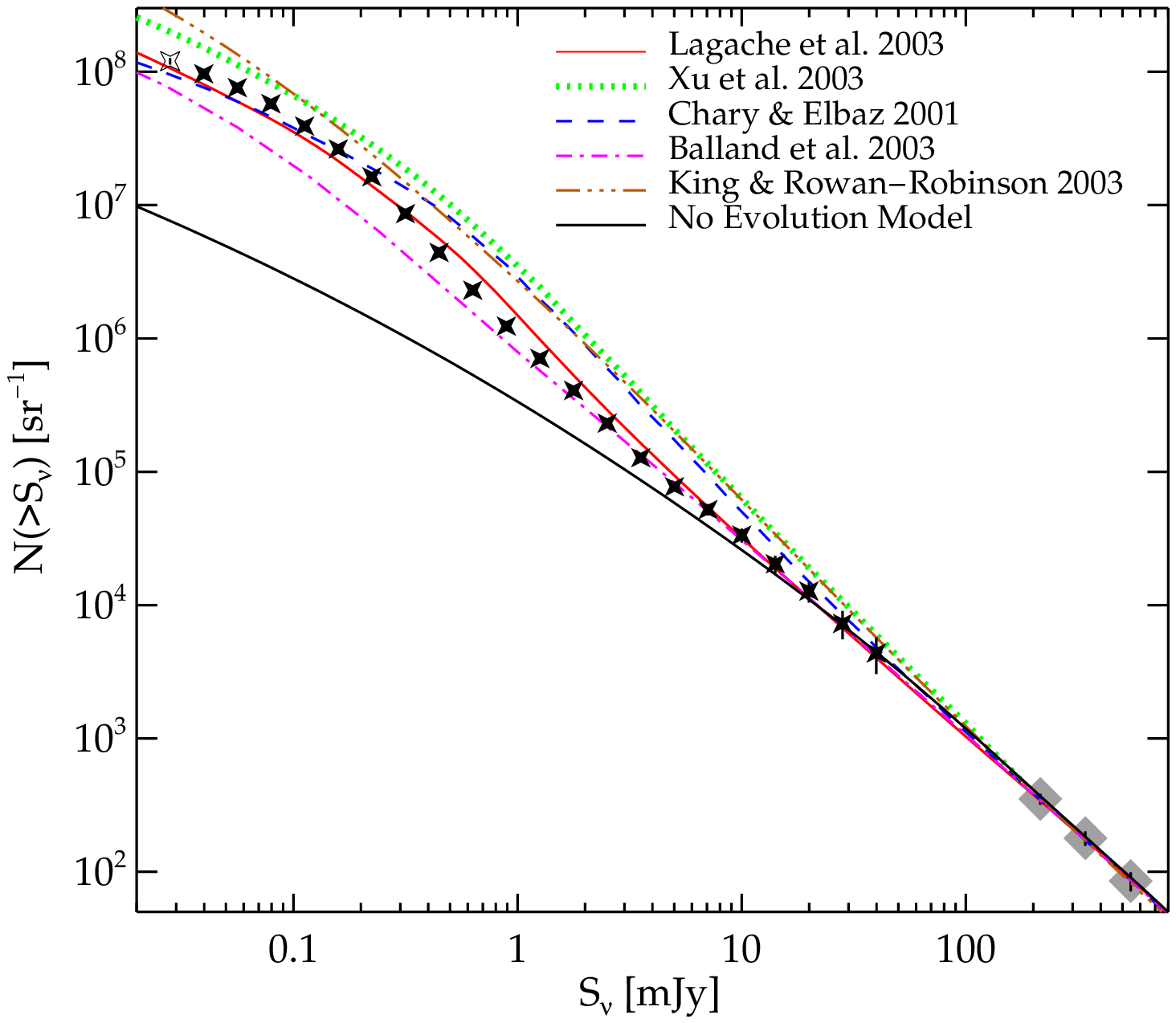}{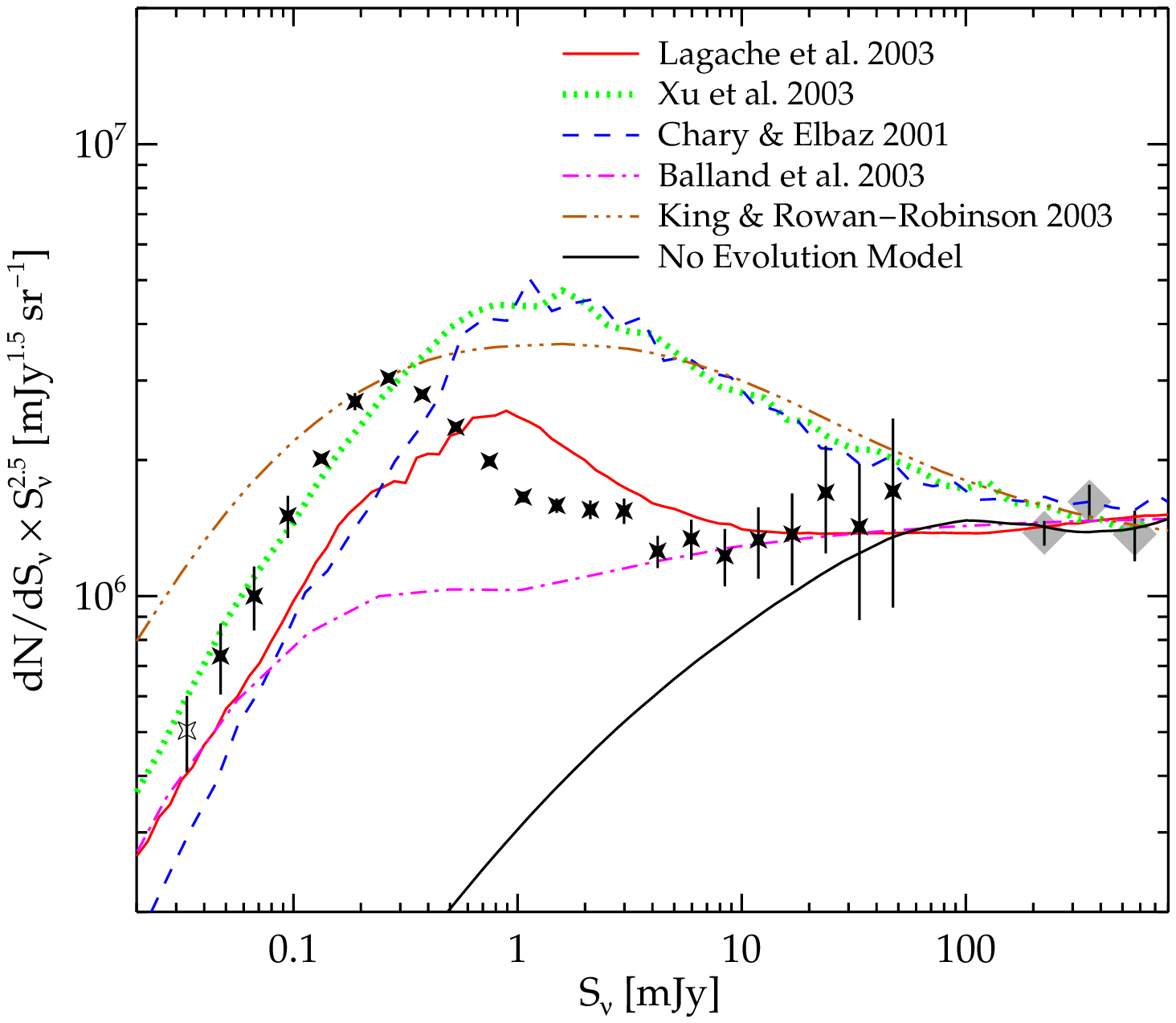}
\caption{ \figtwocap }
\end{figure*}
\fi

To estimate completeness and photometric reliability in the
24~\micron\ source catalogs, we repeatedly inserted artificial sources
into each image with a flux-distribution approximately matching the
measured number counts.  We then repeated the source detection and
photometry process, and compared the resulting photometry to the input
values.  In figure~\ref{fig:completeness}, we show for  the
\textit{Chandra} Deep Field South (CDF--S; one of the deep, \spitzer\ fields)
the relative fraction of sources within $r \leq 2\farcs3$ (half the
FWHM) and with a flux difference $< 50$\% compared to their input
values.  From these simulations, we estimated the flux--density limit
where 80\% of the input sources are recovered with this photometric
accuracy, and these are listed in table~\ref{table:fields}.
Simultaneously, we estimate that down to the 80\% completeness limit
the number of sources that result from fainter sources either by
photometric errors or the merging of real sources is $\lsim 10$\%.
We also repeated the source detection and photometry
process on the negative of each MIPS 24~\micron\ image.  This test
provides an estimate for the number of spurious sources arising from
the noise properties of the image, as shown in
figure~\ref{fig:completeness}.  For all of our fields, the
spurious--source fraction for flux densities greater than the 80\%
completeness limit is $< 10$\%. 


\section{The 24~\micron\ Source Counts}

Figure~\ref{fig:counts} shows the 24~\micron\ cumulative and
differential number counts that have been averaged over the fields listed in
table~\ref{table:fields}. 
The differential and cumulative counts (corrected and uncorrected) are
listed in table~\ref{table:counts}.  
The faintest datum (denoted by the open symbol in figure~\ref{fig:counts})
is derived to the 50\% completeness limit for the European Large--Area
\textit{ISO} Survey (ELAIS) field ($C_{50\%} = 35$~$\mu$Jy).  The
remaining (less deep) fields are used after correcting for
completeness to the 80\% level only.
Error bars in the figure correspond to Poissonian
uncertainties and an estimate for cosmic variance using the standard
deviation of counts between the fields.  For $S_\nu \leq 83$~$\mu$Jy
where the counts are derived solely from the smaller ELAIS field, we
estimate the uncertainty (18\%) using the standard deviation of counts
at faint flux densities in cells of 130~arcmin$^2$ from the CDF--S (see
below).  We have ignored the contribution to the number counts from
stars at 24~\micron, which are negligible  at these
Galactic latitudes and flux densities based on preliminary \spitzer\
observations.

At bright flux densities, $S_\nu \gsim 5$~mJy, the differential
24~\micron\ source counts increase at approximately the Euclidean
rate, $dN/dS_\nu \sim S^{-2.5}$, which extends the trends observed by
the IRAS 25~\micron\ population by two orders of magnitude
\citep{hac91,shu98}.  For $S_\nu \simeq
0.4-4$~mJy, the 24~\micron\ counts increase at super--Euclidean rates,
and peak near $0.2-0.4$~mJy.   This observation is similar to the trend
observed in the \textit{ISO} 15~\micron\ source counts \citep{elb99},
but the peak in the 24~\micron\ differential source counts occurs at
fluxes fainter by a factor $\approx 2.0$.  The peak lies above the
80\%\ completeness limit for nearly all fields, and is seen in the
counts of the fields individually.  Thus the observed turn over is
quite robust.   The counts converge rapidly with a sub--Euclidean rate
at $\lsim 0.2$~mJy.   We broadly fit the faint--end ($S_\nu \simeq
35-130$~$\mu$Jy) of the differential number counts using a powerlaw,
$dN/dS_\nu = C\, (S_\nu/1\,\mathrm{mJy})^{-\alpha}$, with $\alpha =
-1.5 \pm 0.1$.  This result is consistent with a separate analysis
based solely on the ELAIS field \citep{cha04}.

The observed counts are strongly inconsistent with expectations from
non--evolving models of the local IR--luminous population.  In
figure~\ref{fig:counts}, we show the 24~\micron\ counts
derived from the local luminosity function at \textit{ISO} 15~\micron\
\citep{xu00} and assuming local
galaxy SEDs \citep{dal01}.  We have used the the local \textit{ISO}
15~\micron\ luminosity function, because the $k$--correction between
rest--frame \textit{ISO} 15~\micron\ and observed MIPS 24~\micron\
bands is minimized at higher redshifts ($z\sim 0.6$), which is more
appropriate for the counts at fainter flux densities.  However,
using the local \textit{IRAS} 25~\micron\ luminosity function
\citep{shu98} yields essentially identical results.  While the
non--evolving fiducial model is consistent with the observed
24~\micron\ counts for $S_\nu \gsim 20$~mJy, it underpredicts the
counts at $S_\nu \lsim 0.4$~mJy by more than a factor of 10.

Because the deep \spitzer\ fields are measured in many sightlines with
large solid angle, we can estimate the fluctuations in the number of
sources as a function of sky area and flux density in smaller--sized
fields. For example, in the CDF--S, which achieves an
80\%--completeness limit of 83~$\mu$Jy over $\simeq 0.6$~deg$^2$ (see
table~\ref{table:fields} and figure~\ref{fig:completeness}), we have
computed the variance in the number of sources in solid angles of 100
and 300~arcmin$^2$.  For the sources that span the peak in the
differential counts ($0.1-1$~mJy), the fluctuation in the number of
sources in flux--density bins of 0.15~dex is roughly $15$\% in these
areas.  This implies that small--sized fields suffer sizable
field--to--field variation in the number of counts from the cosmic
variance of source clustering. This effect is present in even larger
fields:  the number density of sources brighter than 0.3~mJy  varies
by $\sim 10$\% in fields of $\sim 0.5$~deg$^2$
(table~\ref{table:fields}), and is consistent with fluctuations
expected from galaxy clustering on fields of this size at $z \sim 1$,
(scale lengths of $20-70$~Mpc).    The counts presented here average
over fields from many sightlines and significantly larger areas.  We
conservatively estimate that variations due to galaxy clustering
correspond to uncertainties in the number counts of a less than a few
percent in each flux bin.


\section{Interpretation and Discussion}

The form of the observed 24~\micron\ source counts differs strongly
from predictions of various contemporary models (see
figure~\ref{fig:counts}).  Four of the models are phenomenological in
approach, which parameterize the evolution of IR--luminous galaxies in
terms of density and luminosity to match observed counts from
\textit{ISO}, radio, sub--mm, and other datasets. Several
of these models \citep{cha01,kin03,xu03} show a rapid increase in the
number of sources at super--Euclidean rates at relatively bright flux
densities ($S_\nu \gsim 10$~mJy) and peak near 1~mJy.  These models
generally predict a redshift distribution for the MIPS 24~\micron\
population that peaks near $z\sim 1$, based largely on expectations
from the \textit{ISO} populations, and they  overpredict the
24~\micron\ number counts by factors of $2-3$ at $\sim 1$~mJy.  The
\citet{lag03} model predicts a roughly Euclidean increase in the
counts for $S_\nu > 10$~mJy.  The shape of the counts in this model is
similar to the observed distribution, but it peaks at $S_\nu \sim
1$~mJy, at higher flux densities than the observed counts.  This model
predicts a redshift distribution that peaks near $z\sim 1$, but tapers
slowly with a significant population of IR--luminous galaxies out to
$z\gsim 2$ \citep{dol03}.

The model of Balland, Devriendt, \& Silk (2003) is based on
semi--analytical hierarchical models within the Press--Schecter
formalism in which galaxies identified as `interacting' are assigned
IR--luminous galaxy SEDs.   This model
includes additional physics in that the evolution of galaxies depends
on their local environment and merger/interaction histories.  Although
this model predicts a near--Euclidean increase in the counts for
$S_\nu \gsim 10$~mJy, the counts shift to sub--Euclidean rates at
relatively bright flux densities.  The semi--analytical formalism
seems to not include important physics that are necessary to reproduce
the excess of faint IR sources.  This illustrates the need for
large--area multi--wavelength studies of \spitzer\ sources to connect
optical-- and IR--selected sources at high redshift to understand the
mechanisms that produce IR--luminous stages of galaxy evolution.

The peak in the 24~\micron\ differential number counts occurs at
fainter flux densities than predicted from the phenomenological models
based on the \textit{ISO} results.   This may suggest possibilities
such as a steepening in the slope of the IR luminosity function with
redshift, or evolution in the relation between the mid-- and total
IR. Phenomenological models which reproduce the IR background predict
a faint--end slope of the IR luminosity function that should be quite
shallow at high redshifts, with `\lstar' luminosities that correspond
to $L_{\mathrm{IR}} > 10^{11}$~\lsol\ for $z\gsim 1$ \citep[see][]{hau01}. For
most plausible IR luminosity functions, galaxies with \lstar\
luminosities dominate the integrated luminosity density.
\citet{elb02} observed that the redshift distribution of objects with
these luminosities in deep \textit{ISO} surveys spans  $z\simeq
0.8-1.2$,  and that these objects constitute a large fraction of the
total cosmic IR background.  Therefore, it seems logical that objects
with these luminosities dominate 24~\micron\ number counts at
$0.1-0.4$~mJy, and it follows that their redshift distribution must
lie at $z\sim 1-3$ (i.e., where this flux density corresponds to $\sim
10^{11-12}$~\lsol, using empirical relations from Papovich \& Bell
2002).  Indeed, a similar conclusion is inferred based on a revised
phenomenological model using the 24~\micron\ number counts presented
here \citep{lag04}, and allowing for small changes in the mid--IR SEDs
of IR--luminous galaxies.  Examples of MIPS 24~\micron\ sources at
these redshifts and luminosities have been readily identified in
optical ancillary data \citep{lef04}.  We therefore attribute the
the peak in the 24~\micron\ differential number counts at
fainter flux densities to a population of luminous IR galaxies at
redshifts higher than explored by \textit{ISO}.

Integrating the differential source--count distribution provides an
estimate for their contribution to the cosmic IR background at
24~\micron, i.e, $I_\nu = \int dN/dS_\nu\, S_\nu\, dS_\nu$.    For
sources brighter than 60~$\mu$Jy, we derive a lower limit on the
total background of $\nu I_\nu(24\micron) = 1.9\pm0.6$~nW m$^{-2}$
sr$^{-1}$.   Due to the steep nature of the source counts, most of
this background emission results from galaxies with fainter apparent
flux densities.  We find that $\simeq 60$\% of the 24~\micron\
background originates in galaxies with $S_\nu \leq 0.4$~mJy, and therefore
the galaxies responsible for the peak in the differential source
counts also dominate the total background emission.   Our result is
consistent with the \textit{COBE} DIRBE upper limit
$\nu\,I_\nu(25\micron) < 7$~nW m$^{-2}$ sr$^{-1}$ inferred from
fluctuations in the IR background \citep{kas00,hau01}.
As a further estimate on the total 24~\micron\ background
intensity, we have extrapolated the number counts for $S_\nu <
60$~$\mu$Jy using the fit to the faint--end slope of the 24~\micron\
number counts in \S~3.   Under this assumption, we find that
sources with $S_\nu < 60$~$\mu$Jy would contribute
$0.8^{+0.9}_{-0.4}$~nW m$^{-2}$ sr$^{-1}$ to the 24~\micron\
background, which when summed with the above measurement yields an
estimate of the total background of $\nu
I_\nu^{(\mathrm{tot})}(24\micron) = 2.7^{+1.1}_{-0.7}$~nW m$^{-2}$
sr$^{-1}$.  For this value, the sources detected in the
deep \spitzer\ 24~\micron\ surveys produce $\sim 70$\% of the total
24~\micron\ background.

\acknowledgements 

We acknowledge our colleagues for stimulating conversations, the SSC
staff for efficient data processing, Thomas Soifer and
the IRS team for executing the Bo\"otes--field observations, Daniel Eisenstein
for cosmology discussions, Jim Cadien for his
assistance with the data reduction, and the entire
\spitzer\ team for their concerted effort.   We also thank the
referee, Matthew Malkan, for a thorough and insightful
report.  Support for this work was provided by NASA through Contract
Number 960785 issued by JPL/Caltech.



\ifsubmode
\newpage
\else
\fi

\ifsubmode
\begin{deluxetable}{lcccccccc}
\rotate
\tableone
\end{deluxetable}
\fi

\ifsubmode
\begin{deluxetable}{cccccccccc}
\tabletwo
\end{deluxetable}
\else
\begin{deluxetable}{cccccccccc}
\tabletwo
\end{deluxetable}
\fi

\ifsubmode
\begin{figure}
\plotone{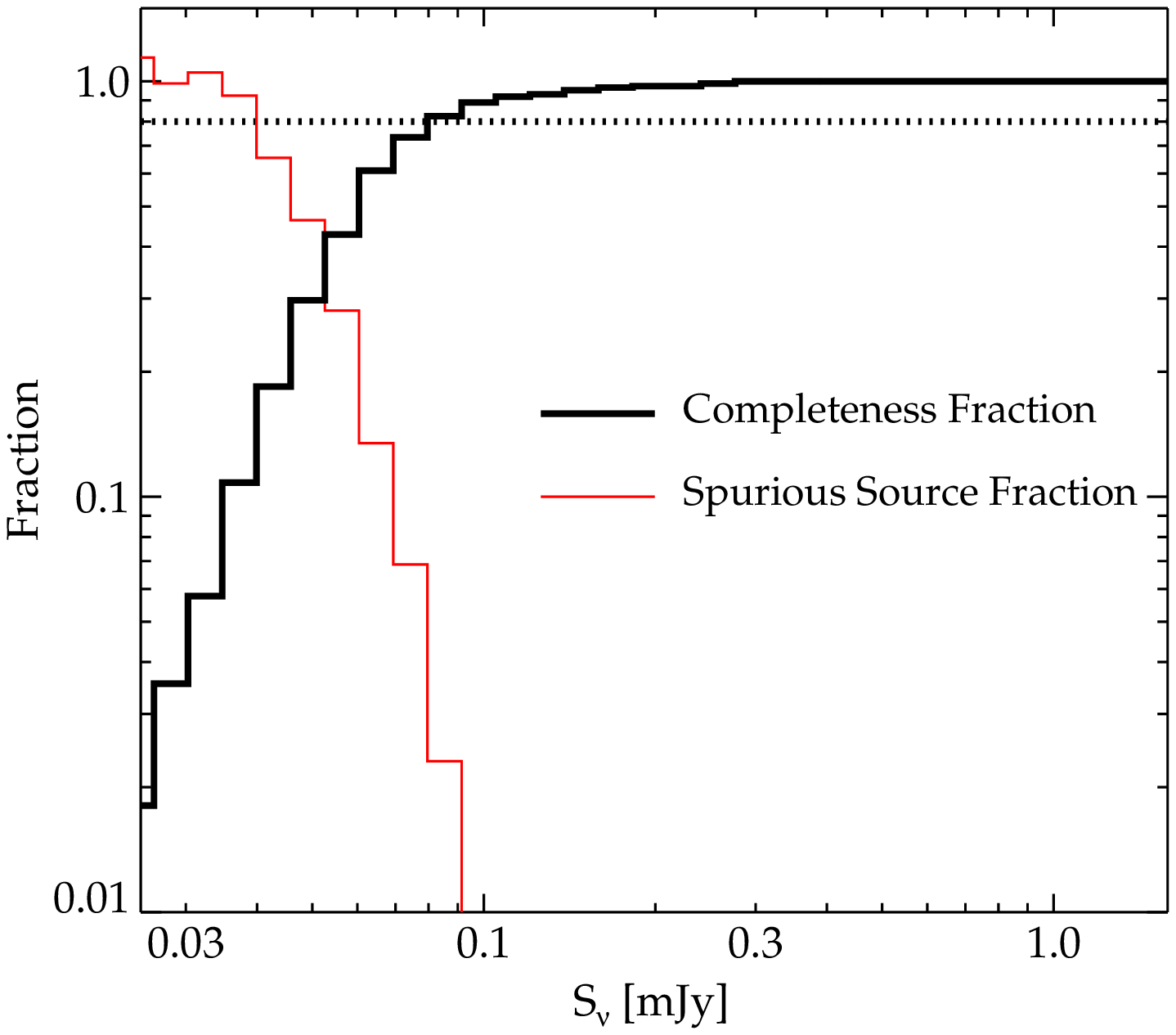}
\caption{ \figonecap }
\end{figure}

\begin{figure}
\epsscale{1.111}
\plottwo{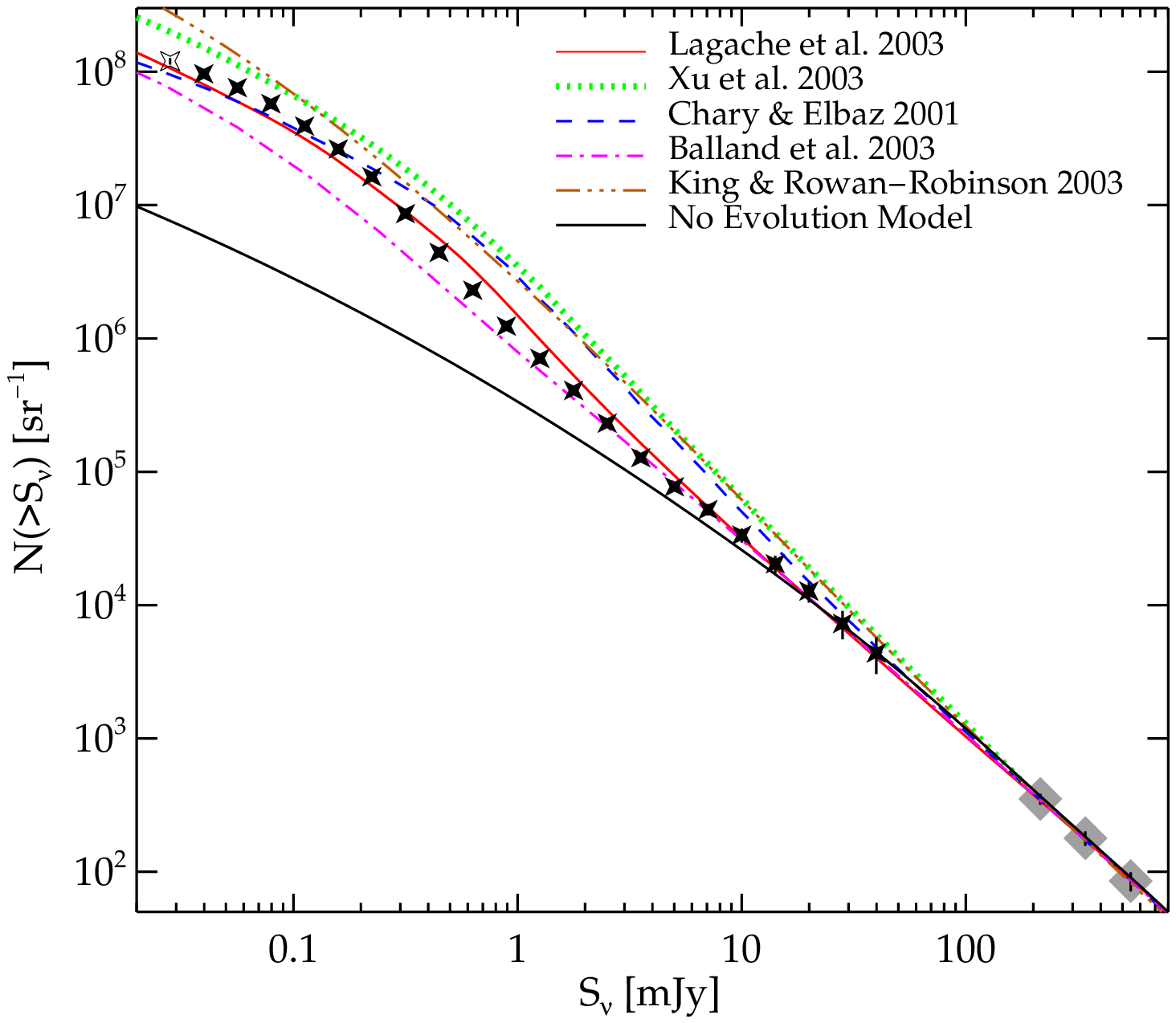}{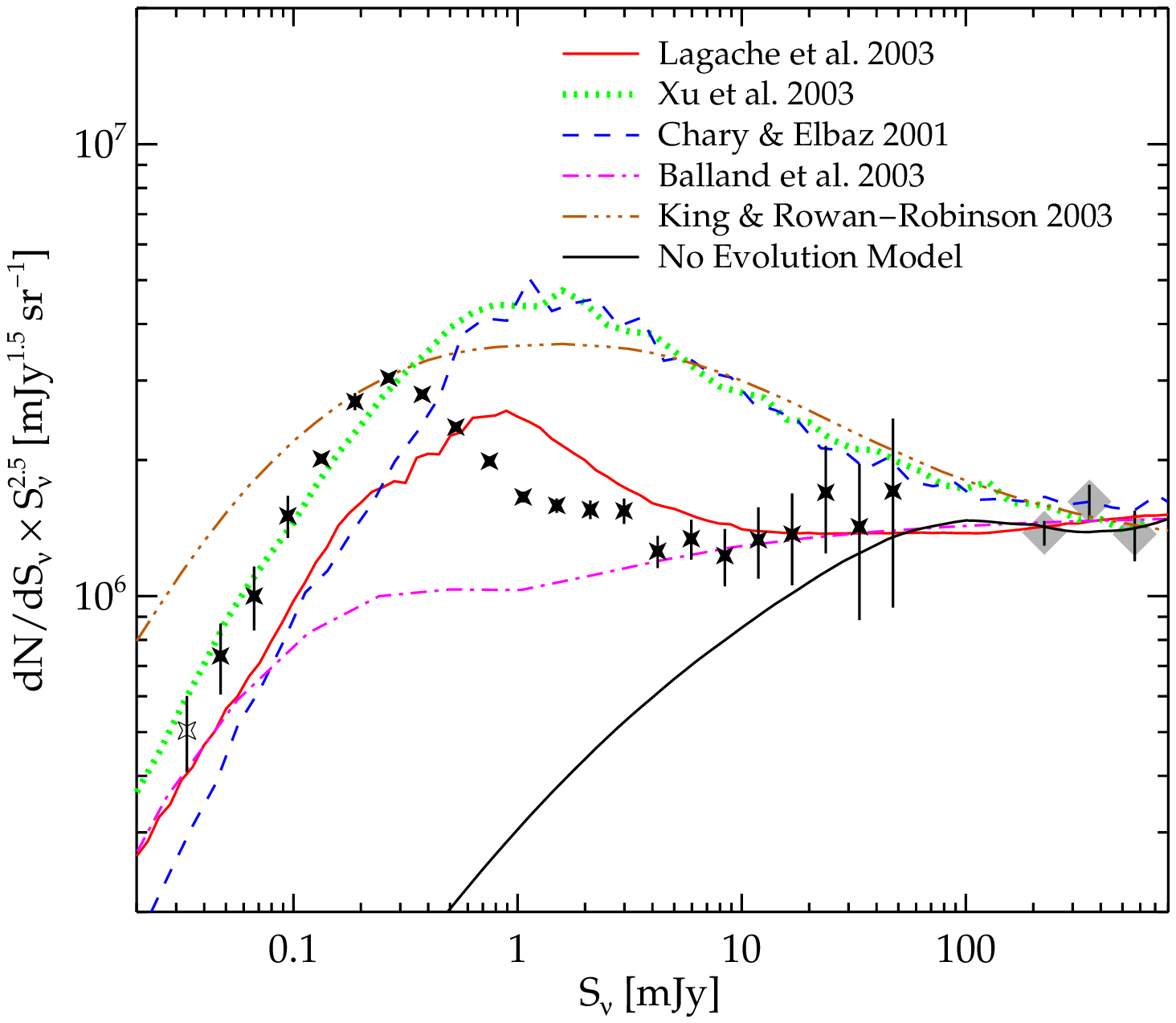}
\caption{ \figtwocap }
\end{figure}
\fi
\end{document}